\documentclass[aps,pre,twocolumn,superscriptaddress,floatfix]{revtex4-1}

\usepackage{graphicx,bm,amssymb,amsmath,dcolumn,hyperref}
\usepackage{multirow}
\usepackage{enumerate}
\usepackage{enumitem}
\usepackage{color}
\usepackage{subfigure}  
\usepackage{epstopdf}
\usepackage{bbm}
\usepackage{graphicx}
\usepackage{hyperref}
\usepackage{threeparttable}
\usepackage{amsthm}
\usepackage{mathtools}
\usepackage{color}
\usepackage{tikz}
\usepackage{float}

\begin{document}
\newcommand{\upcite}[1]{\textsuperscript{\textsuperscript{\cite{#1}}}}
\newcommand{\be}{\begin{equation}}
\newcommand{\ee}{\end{equation}}
\newcommand{\half}{\frac{1}{2}}
\newcommand{\ith}{^{(i)}}
\newcommand{\im}{^{(i-1)}}
\newcommand{\gae}
{\,\hbox{\lower0.5ex\hbox{$\sim$}\llap{\raise0.5ex\hbox{$>$}}}\,}
\newcommand{\lae}
{\,\hbox{\lower0.5ex\hbox{$\sim$}\llap{\raise0.5ex\hbox{$<$}}}\,}

\definecolor{blue}{rgb}{0,0,1}
\definecolor{red}{rgb}{1,0,0}
\definecolor{green}{rgb}{0,1,0}
\newcommand{\blue}[1]{\textcolor{blue}{#1}}
\newcommand{\red}[1]{\textcolor{red}{#1}}
\newcommand{\green}[1]{\textcolor{green}{#1}}
\newcommand{\orange}[1]{\textcolor{orange}{#1}}
\newcommand{\yd}[1]{\textcolor{blue}{#1}}

\newcommand{\scrA}{{\mathcal A}}
\newcommand{\scrE}{{\mathcal E}} 
\newcommand{\scrF}{{\mathcal F}} 
\newcommand{\scrL}{{\mathcal L}}
\newcommand{\scrM}{{\mathcal M}} 
\newcommand{\scrN}{{\mathcal N}}
\newcommand{\scrS}{{\mathcal S}}
\newcommand{\scrs}{{\mathcal s}}
\newcommand{\scrP}{{\mathcal P}}
\newcommand{\scrO}{{\mathcal O}}
\newcommand{\scrR}{{\mathcal R}}
\newcommand{\scrC}{{\mathcal C}}
\newcommand{\scrV}{{\mathcal V}}
\newcommand{\scrD}{{\mathcal D}}
\newcommand{\scrG}{{\mathcal G}}
\newcommand{\scrW}{{\mathcal W}}
\newcommand{\PP}{\mathbb{P}}
\newcommand{\ZZ}{\mathbb{Z}}
\newcommand{\EE}{\mathbb{E}}
\renewcommand{\d}{\mathrm{d}}
\newcommand{\dm}{d_{\rm min}}
\newcommand{\rhojunction}{\rho_{\rm j}}
\newcommand{\rhojunctionLim}{\rho_{{\rm j},0}}
\newcommand{\rhobranch}{\rho_{\rm b}}
\newcommand{\rhobranchLim}{\rho_{{\rm b},0}}
\newcommand{\rhononbridge}{\rho_{\rm n}}
\newcommand{\rhononbridgeLim}{\rho_{{\rm n},0}}
\newcommand{\percolationCluster}{C}
\newcommand{\leafFreeCluster}{C_{\rm \ell f}}
\newcommand{\bridgeFreeCluster}{C_{\rm bf}}
\newcommand{\df}{d_{\rm f}}
\newcommand{\yt}{y_{\rm t}}
\newcommand{\yh}{y_{\rm h}}
\newcommand{\dfprime}{d'_{\rm f}}
\newcommand{\yhhat}{\hat{y}_{\rm h}}
\newcommand{\ythat}{\hat{y}_{\rm t}}
\newcommand{\yhstar}{y^*_{\rm h}}
\newcommand{\ytstar}{y^*_{\rm t}}
\newcommand{\zc}{z_{\rm c}}
\newcommand{\dc}{d_{\rm c}}
\newcommand{\bfx}{{\bf x}}
\newcommand{\bfO}{{\bf o}}
\newcommand{\bfo}{{\bf o}}
\newcommand{\bfS}{{\bf S}}
\newcommand{\bfr}{\bf r}
\newcommand{\origin}{\bf 0}
\newcommand{\bfe}{\bf e}
\newcommand{\bfk}{{\bf k}}
\newcommand{\bfy}{\bf y}
\newcommand{\bfu}{\bf u}
\newcommand{\bmomega}{{\bm \omega}}
\newcommand{\bfU}{{\bf u}}
\newcommand{\ind}{\mathbbm{1}}
\newcommand{\xiu}{\xi_{\rm u}}

\title{Logarithmic finite-size scaling of the self-avoiding walk at four dimensions}
\date{\today}
\author{Sheng Fang}
\affiliation{MinJiang Collaborative Center for Theoretical Physics,
	College of Physics and Electronic Information Engineering, Minjiang University, Fuzhou 350108, China}
\affiliation{Hefei National Laboratory for Physical Sciences at Microscale and 
	Department of Modern Physics, University of Science and Technology of China, 
	Hefei, Anhui 230026, China}
\author{Youjin Deng}
\email{yjdeng@ustc.edu.cn}
\affiliation{MinJiang Collaborative Center for Theoretical Physics,
	College of Physics and Electronic Information Engineering, Minjiang University, Fuzhou 350108, China}
\affiliation{Hefei National Laboratory for Physical Sciences at Microscale and 
	Department of Modern Physics, University of Science and Technology of China, 
	Hefei, Anhui 230026, China}

\author{Zongzheng Zhou}
\email{eric.zhou@monash.edu}
\affiliation{ARC Centre of Excellence for Mathematical and Statistical Frontiers (ACEMS),
	School of Mathematics, Monash University, Clayton, Victoria 3800, Australia}

\begin{abstract}

The $n$-vector spin model, which includes the self-avoiding walk (SAW) as a special case 
for the $n \rightarrow 0 $ limit, has an upper critical dimensionality at four spatial dimensions (4D). 
We simulate the SAW on 4D hypercubic lattices 
with periodic boundary conditions by an irreversible Berretti-Sokal algorithm 
up to linear size $L=768$. From an unwrapped end-to-end distance, 
we obtain the critical fugacity as $\zc= 0.147 \, 622 \, 380(2)$, 
improving over the existing result $\zc=0.147 \, 622 \, 3(1)$ by  50 times. 
Such a precisely estimated critical point enables us to perform a systematic study of 
the finite-size scaling of 4D SAW for various quantities. 
Our data indicate that near $\zc$, the scaling behavior of the free energy 
simultaneously contains a scaling term from the Gaussian fixed-point and the other 
accounting for multiplicative logarithmic corrections. 
In particular, it is clearly observed that the critical magnetic susceptibility 
and the specific heat logarithmically diverge 
as $\chi \sim L^2 (\ln L)^{2 \yhhat}$ and $C \sim (\ln L)^{2 \ythat}$, 
and the logarithmic exponents are determined  as $\yhhat=0.251(2)$ and $\ythat=0.25(3)$, 
in excellent agreement with the field theoretical prediction $\yhhat=\ythat=1/4$.  
Our results  provide a strong support for the recently conjectured finite-size scaling form 
for the O$(n)$ universality classes at 4D.

\end{abstract}
\pacs{05.50.+q (lattice theory and statistics), 05.70.Jk (critical point phenomena),
64.60.F- (equilibrium properties near critical points, critical exponents)}
\maketitle

\section{Introduction}
\label{Introduction}
The self-avoiding walk (SAW) is a special random walk without self-intersection, 
and plays an important role in combinatorics, probability, polymers sciences 
and statistical mechanics~\cite{madras2013self}. 
The SAW on a connected graph is simply a sequence of distinct vertices 
and each consecutive pair of vertices are adjacent. 
Most studies about SAWs focus on $d$-dimensional hypercubic lattices, 
where walks start from the origin. 
Let $c_N$ denote the number of $N$-step SAWs,  the generating function is
	\begin{equation}
	\label{equ:chi_z}
	\chi(z) = \sum_{N=0}^{\infty} c_N z^N = \sum_{\omega} z^{|\omega|},
	\end{equation} 
where $\sum_{\omega}$ is  over all walks and $|\omega|$ is the walk length. 
The parameter $z$ ($z>0$) is called the fugacity or activity. 
The summation converges for $z<\zc$, where the critical value $\zc$ is the convergence radius.
Similarly, the two-point function with the end point at coordinate $\bfx$ is defined as
\begin{equation}
\label{equ:define_gx}
G(\bfx) =\sum_{N=0}^{\infty} c_N(\bfx) z^N =  \sum_{\omega:0\rightarrow \bfx} z^{|\omega|},
\end{equation}
where $c_N(\bfx)$ denotes the number of $N$-step SAWs ending at $\bfx$. 
Obviously $\chi(z)=\sum_{\bfx} G(\bfx)$, so the generating function $\chi(z)$ is 
simply the susceptibility of SAW. 
For the ensemble of SAWs with fixed length $N$, called the canonical ensemble, 
the mean displacement is $\xi_N = \sqrt{\frac{1}{c_N} \sum_{\omega:|\omega|=N} 
\|\bfx_{\omega}\|^2}$  with $\bfx_{\omega}$  the endpoint of the walk and it is expected that
	\begin{gather}
	\label{equ:cn}
		c_N \sim  \mu^N N^{\gamma -1}\;, \ \ \xi_N  \sim   N^{\nu}\;.
	\end{gather}
The connectivity constant $\mu$ is related to the critical fugacity as $\mu = 1/\zc$, 
and $\gamma, \nu$ are critical exponents. 
Since $\xi_N$ plays a role of correlation length, $\nu$ is normally referred to be 
the correlation-length exponent.
In the ensemble of SAWs with fixed fugacity $z$, called the grand-canonical ensemble, 
the mean displacement is $\xi(z) = \sqrt{\sum_{\omega} z^{|\omega|} 
\|\bfx_{\omega}\|^2/\sum_{\omega} z^{|\omega|}}$. 
When $z$ approaches $\zc$ from below, it is expected that  
        \begin{gather}
            \chi(z)  \sim (\zc - z)^{-\gamma}\;, \ \  \xi(z) \sim (\zc - z)^{-\nu}\;,
     \end{gather}
and $G(\bfx)   \sim \|\bfx\|^{2-d+\eta}$ at $\zc$.

Over the past few decades, the SAW has been extensively studied. In two dimensions (2D),
the Coulomb-gas theory predicted the critical exponents  $\gamma = 43/32$ 
and $\nu = 3/4$~\cite{nienhuis1982exact}. 
This can even be established rigorously by proving that the scaling limit of SAW 
is ${\rm SLE}_{8/3}$ (Schramm-Loewner evolution)~\cite{lawler2002scaling}, 
conditioned on the conjecture that the scaling limit of SAW exists and is conformally invariant. 
In 2D, the exact value of critical point is only known on the honeycomb lattice, 
which was first conjectured as $z_c = 1/\sqrt{2+\sqrt{2}}$ in 1982~\cite{nienhuis1982exact} 
and after 30 years was finally proved rigorously~\cite{duminil2012connective}. 
Numerical estimates of critical points on other 2D lattices are summarized in Table~\ref{Tab:Num_saw}. 
In particular, the value of $\zc$ on the square lattice was estimated to the 15th decimal place, 
an unprecedent precision, using the topological transfer-matrix method~\cite{jacobsen2016growth}. 
No rigorous results are available in 3D. Estimates of $\zc$ on various 3D lattices 
are shown in Table~\ref{Tab:Num_saw}, and the critical exponents were determined 
to a very high precision as  $\gamma = 1.156\,953\,00(95)$ \cite{clisby2007self} 
and $\nu=0.587\,597\,00(40)$   \cite{clisby2016high}. 
For $d\geq 5$, it was proved via lace expansion~\cite{hara1992self,hara2008decay} 
that the two-point function $G(\bfx) \sim \|\bfx\|^{2-d}$ and 
critical exponents $\gamma = 1$, $\nu =1/2$, consistent with 
the prediction of the Gaussian fixed point in the framework of renormalization group
(frequently referred to be the mean-field theory).
Critical points on high-dimensional hypercubic lattices were estimated 
in Refs.~\cite{owczarek2001scaling,hu2017irreversible}, summarized in Table~\ref{Tab:Num_saw}.

For SAW, $\dc =4$ is the upper critical dimension, 
where scaling behavior can be described by mean-field theory but with multiplicative logarithmic correction. 
In Refs.~\cite{Duplantier1987Polymer,grassberger1994self}, renormalization-group arguments give that
 	\begin{equation}
	\label{equ:cn1}
	c_N \sim  \mu^N [\ln(N/N_0)]^{1/4}\;, \ \ \xi_N  \sim   N^{1/2} [\ln(N/N_0)]^{1/4}\;,
	\end{equation}
with some positive constant $N_0$. However, to numerically confirm the logarithmic scaling is a big challenge, 
since one has to simulate very long self-avoiding walks. 
Not until recently, Clisby~\cite{clisby2018monte} simulated SAWs with walk length up to one billion steps 
using an improved pivot algorithm~\cite{clisby2010accurate,clisby2010efficient}, 
and finally the logarithmic scaling was clearly observed for the first time. 
By universality arguments, the behavior of $\chi(z)$ and $G(\bfx)$ for 4D SAW 
is believed to exhibit the same scaling as the weakly self-avoiding walk; 
for the latter it has been rigorously proved that~\cite{bauerschmidt2015logarithmic,bauerschmidt2015critical}
\begin{eqnarray}
\label{equ:4dSAW_chi}
&G(\bfx) \sim \|\bfx\|^{-2}\;, \nonumber \\
&\chi(z) \sim (\zc-z)^{-1} [-\ln(\zc-z)  ]^{1/4}\;.
\end{eqnarray}

A well-known connection between SAW and spin systems, pointed out by de Gennes in 1970's, is that SAW can be interpreted as the $n=0$ case of the $n$-vector (or O($n$)) model. The O$(n)$ model is to describe a system of interactive spins, and its Hamiltonian is
\begin{equation}
H = - \sum_{\bfx\sim \bfy} \bfS_{\bfx} \cdot \bfS_{\bfy},
\end{equation}
where the summation is over all pairs of adjacent vertices, and the spins $\bfS_{\bfx}$ are $n$-component unit vectors. Special cases of the O$(n)$ model are the Ising model ($n=1$), the XY model ($n=2$) and the Heisenberg model ($n=3$). In particular, under the graphical (loop) representation, it can be shown that SAW can be interpreted as a $0$-component spin model, via the fact that the two-point function of the $n$-vector model reduces to that of SAW as $n\rightarrow 0$~\cite{madras2013self}. It is known that the upper critical dimension is $\dc = 4$ for the general O$(n)$ model, and is believed that for $d >4$, the $n$-vector model exhibits the same critical behavior for all $n\geq 0$.

Recently, it receives much attention to study the scaling behavior of SAW on high-dimensional boxes~\cite{GrimmElciZhouGaroniDeng2017,ZhouGrimmFangDengGaroni2018,DengGaroniGrimmZhou21},  for understanding the \emph{finite-size scaling} (FSS) of the general $n$-vector model for $d > \dc$~\cite{FernandezFrohlichSokal13}, which has been the subject of long-standing debate since 1980s. For the convenience of readers, we shall briefly recall some basic aspects of the FSS theory in critical phenomena, and a brief review of the subtleties and recent development of FSS in high dimensions.

FSS is a fundamental theory to describe the asymptotic approach of finite systems to thermodynamic limit near a continuous phase transition point. The main conjecture of FSS is that the correlation length is cut off by the order of linear system size, 
and the singular part of the free energy function can be written as 
            \begin{equation}
             \label{equ:free_energy}
                f(t,h) = L^{-d}\tilde{f}(tL^{\yt},hL^{\yh}),
            \end{equation}
where  the parameters $t$, $h$ represent the thermal and magnetic scaling fields, 
and $\yt, \yh$ are the corresponding renormalization group exponents with $\yt = 1/\nu$ , $\yh = d - \beta/\nu$. Here $\beta$ and $\nu$ are the critical exponents for order parameter and correlation length, respectively. Below upper critical dimensions $\dc$, FSS has been widely accepted, and proved to be a powerful tool to extract critical points and exponents from finite-size systems.

\begin{table}[ht!]
 	\begin{tabular}{ | p{0.5cm}| p{3.8cm} p{3.8cm}|} 
 		\hline
 		$d$ & \centering{$\zc$} & \\ 
 		\hline
 		  & $1/{\sqrt{2+\sqrt{2}}}$ \upcite{nienhuis1982exact,duminil2012connective} &(honeycomb)  \\
 		  &	0.379\,052\,277\,755\,161(5) \upcite{jacobsen2016growth} &(square, TTM) \\
 		2 & 0.240\,917\,574\,5(15)\upcite{jensen2004self}&(triangle, EM )  	  \\		
 		  & 0.390\,537\,012\,5(15)\upcite{jensen2004improved}    &(kagom{\'e}, EM) \\	
 		\hline
 		  & 0.213\,491\,0(3)\upcite{hsu2004polymers} &(simple cubic, MC)  \\
 		3 &0.099\,630\,6(2)\upcite{schram2017exact}  &(face-centered cubic, MC) \\
  		  &0.153\,127\,2(5)\upcite{schram2017exact}  &(body-centered cubic, MC)  \\		    
 		\hline
 		4 &0.147\,622\,3(1)\upcite{owczarek2001scaling} &   \\
 		  &0.147\,622\,380(2)(this work)     & \\ 
 		\cline{1-3} 
 		5 &0.113\,140\,843(5)\upcite{hu2017irreversible}  & \\
 		\cline{1-3} 
 		6 & 0.091\,927\,87(3)\upcite{owczarek2001scaling} & \\
 		\cline{1-3} 
 		7 & 0.077\,502\,46(2)\upcite{owczarek2001scaling} &\\
 		\cline{1-3} 
 		8 & 0.067\,027\,467(9)\upcite{owczarek2001scaling} & \\
 		\hline
 	\end{tabular}
 \caption{Summary of critical points on various lattices and dimensions. The type of lattices and methods are shown in brackets. Here TTM, EM, and MC are short for topological transfer-matrix method, exact enumeration and Monte Carlo methods, respectively. Estimate of $\zc$ for $d\ge 4$ are only available on hypercubic lattices using Monte Carlo methods.}
 \label{Tab:Num_saw}
\end{table}

 However, the FSS above the upper critical dimension is surprisingly subtle. For $d>\dc$, the critical behavior is controlled by the Gaussian fixed point~\cite{ma1976modern}, where one has $\yt = 2$ and $\yh = 1+d/2$. So, together with the FSS ansatz, it predicts that at the critical point the susceptibility scales as $\chi \sim L^{2\yh -d} = L^2$ on high-dimensional boxes where $L$ is the linear system size. Although this is well accepted on hypercubic lattices with free boundary conditions and consistent with numerical observations later on~\cite{LundowMarkstrom2011,LundowMarkstrom2014,WittmannYoung2014,
 FloresSolaBercheKennaWeigel2016}, subtleties appear when one considers the case of periodic boundary conditions(PBC). In 1982, Br\'ezin~\cite{Brezin1982} theoretically showed that the Gaussian-fixed-point prediction fails on hypercubic lattices with PBC. In the next year, this failure was repaired by Binder and Privman~\cite{Fisher1983,PrivmanBinder1983,Binder1987} using the mechanism of \emph{dangerous irrelevant variables}, the effect of which leads to two new renormalized exponents $\ytstar = d/2, \yhstar = 3d/4$. As a consequence, the susceptibility was predicted to scale as $\chi \sim L^{2\yhstar -d} = L^{d/2}$ for the PBC case, distinct with the Gaussian fixed-point prediction $\chi \sim L^2$. In 1985, Binder \emph{et al.}~\cite{BinderNauenbergPrivmanYoung1985,Binder1985} showed some numerical evidence to the scaling $\chi \sim L^{d/2}$ by simulating the Ising model on the 5D hypercubic lattices with PBC with system sizes up to $L=7$. In the same year, by studying the $\phi^4$ field Hamiltonian, Br\'ezin and Zinn-Justin~\cite{BrezinJustin1985} theoretically confirmed the scaling of $\chi \sim L^{d/2}$. Other Monte Carlo studies providing consistent scaling of susceptibility  can be found in Refs.~\cite{LuijtenBlote1997,Mon1996,ParisiLorenzo1996,BloteLuijten1997,
 LuijtenBinderBlote1999,BinderLuijtenMullerWildingBlote1999,
 LundowMarkstrom2011,BercheKennaWalter2012}.

To understand the FSS of $\chi$ above $\dc$ and with PBC, one should start with the more fundamental quantity, the two-point function $G(\bfx)$, since $\chi =\sum_{\bfx}G(\bfx)$. In 1997, by studying the long-range Ising model in one dimension with PBC, Luijten and Bl\"ote~\cite{LuijtenBlote1997} numerically observed that $G(\bfx)$ behaves differently for short and long distances above $\dc$. For the Ising model on the 5D torus, it was claimed in Ref.~\cite{luijten1997interaction} that $G(\bfx)$ still behaves Gaussian-like $\|\bfx\|^{-3}$ for the short distance, but then becomes $\|\bfx\|^{-5/2}$ for the large distance. In 2006, Papathanakos~\citep{Papathanakos2006} conjectured that for the Ising model on high-dimensional ($d>4$) tori, the two-point function
\begin{equation}
 	        \label{equ:5DSAW_finite_gx}
	        G(\bfx) \sim 
	            \begin{cases}
	                \|\bfx\|^{2-d} &\text{if}~\|\bfx\| \leq O\left(L^{d/[2(d-2)]} \right)\\ 
	                L^{-d/2}  &\text{if}~\|\bfx\| \geq O\left(L^{d/[2(d-2)]}\right)\;,
	            \end{cases}
        \end{equation}
and proved that the right-hand side of Eq.~\eqref{equ:5DSAW_finite_gx} is a lower bound of $G(\bfx)$. This conjecture means that $G(\bfx)$ still follows the Gaussian fixed-point prediction for small $\|\bfx\|$, but enters a distance-independent plateau $L^{-d/2}$ for large $\|\bfx\|$. According to Eq.~\eqref{equ:5DSAW_finite_gx}, the scalng of the susceptibility, $\chi \sim L^{d/2}$, comes from the plateau, instead of the algebraic decaying of $G(\bfx)$. On the other hand, Kenna and Berche proposed \cite{KennaBerche2014} in 2014 a different scenario that above $\dc$, the exponent $\eta$ governing the behavior of the two-point function should be replaced by a new exponent $\eta_{\rm Q}$, i.e., $G(\bfx) \sim \|\bfx\|^{2-d-\eta_{\rm Q}}$ with $\eta_{\rm Q} = 2- d/2$. This disagrees with the Gaussian fixed-point prediction, but is consistent with the numerical observation $G(L/2) \sim L^{-d/2}$ and $\chi \sim L^{d/2}$. In 2014, by studying the Fourier modes of susceptibility $\chi_{\bfk}$ for the Ising model on the 5D torus, Young and Wittmann numerically observed that $\chi_{\bfk \neq 0} \sim L^2$ and  $\chi_{\bfk = 0} \sim L^{d/2}$. Namely, the nonzero mode $\chi_{\bfk \neq 0}$ still follows the Gaussian fixed-point prediction, which clearly refuted the proposed scaling of $G(\bfx)$ involving $\eta_{\rm Q}$, but the zero mode $\chi_{\bfk = 0}$ shows the anomalous scaling $L^{d/2}$. Later on, Grimm {\it et al}~\cite{GrimmElciZhouGaroniDeng2017,ZhouGrimmFangDengGaroni2018,
DengGaroniGrimmZhou21} studied the two-point function of SAW and the Ising model on five-dimensional tori, which, by universality arguments, should exhibit the same scaling behavior. By simulating system sizes up to $L=221$ for SAW and $L=101$ for the Ising model, they provided strong numerical evidence to Eq.~\eqref{equ:5DSAW_finite_gx}.

It is believed that the complete graph can be used to model the infinite-dimensional lattices with PBC. Thus, we argue that the distance-independent regime of the two-point function in Eq.~\eqref{equ:5DSAW_finite_gx} can be regarded as the contribution from the complete graph. Indeed, one can show that on the complete graph the correlation between any two vertices scales as $V^{-1/2}$, which corresponds to $L^{-d/2}$ on the lattices once we match the volume $V=L^d$. The distance-dependent behavior in Eq.~\eqref{equ:5DSAW_finite_gx} still follows the Gaussian fixed-point prediction.
So, it suggests that, in order to completely describe the FSS behavior of the Ising model above $\dc$ with PBC, we need both the Gaussian fixed-point asymptotics and the complete-graph asymptotics and the FSS of the free energy density should be written as
      \begin{equation}
        \label{equ:freeenergy_5d}
            f(t,h) = L^{-d} \tilde{f_0}(tL^{\yt},hL^{\yh}) + L^{-d} \tilde{f_1}(tL^{\ytstar},hL^{\yhstar})\;.
        \end{equation}
From universality, it is expected that Eq.~\eqref{equ:freeenergy_5d} applies to the general O$(n)$ model above $\dc=4$. The term $\tilde{f}_0$ arises from the Gaussion fixed point with $(\yt,\yh) = (2, 1+d/2)$,
and accounts for the FSS of distance-dependent observables including the correlation function $G(\bfx)$ 
for short $\|\bfx\|$ and the nonzero Fourier mode $\chi_{\bfk \neq 0}$ etc.
The term $\tilde{f}_1$ can be regarded to correspond to the complete-graph asymptotics with $(\ytstar, \yhstar) = (d/2,3d/4)$, 
and acts as the ``background" contribution describing the leading FSS of the conventional observables,
which include the magnetization, the energy, the magnetic susceptibility and the specific heat etc. In particular, $\tilde{f}_1$ describes the $L$-dependence of the plateau of $G(\bfx)$ for large $\|\bfx\|$, which leads to the scaling $\chi \sim L^{d/2}$. This is consistent with the asymptotics of the susceptibility of the Ising model on the \emph{complete graph}\footnote{A complete graph with $V$ vertices is a graph in which each vertex is connected to all others.} with $V$ vertices, $\chi \sim V^{1/2}$ \cite{luijten1997interaction}, by setting $V=L^d$. Moreover, a dimensionless ratio, defined from the SAW walk length and its fluctuation, was observed to quickly converge to the value analytically obtained from the complete-graph SAW~\cite{DengGaroniGrimmZhou21}, giving a strong support to the conjecture that the term $\tilde{f}_1$ corresponds to the complete-graph asymptotics.

Rich geometric phenomena, consistent with Eq.~(\ref{equ:5DSAW_finite_gx}) and \eqref{equ:freeenergy_5d}, are also observed.
By studying the Fortuin-Kasteleyn clusters of the 5D Ising model, 
the authors in Ref.~\cite{FangGrimmZhouDeng2020} further demonstrated that 
the largest cluster follows the complete-graph asymptotic and other clusters follow the Gaussian fixed-point prediction. Therefore, the Gaussian fixed-point term $\tilde{f}_0$ in Eq.~\eqref{equ:freeenergy_5d} not only dominates the scaling behavior of many Fourier-transformed quantities with nonzero modes, but also determine the behavior of quantities with scale much smaller than the system size.
For the 5D SAW and the loop representation of the 5D Ising model, 
to account for how many times the walk (loop) wraps around the periodic boundaries,
an unwrapped distance $\bfU$ was introduced such that, effectively, the finite box is 
reciprocally placed in the infinite space with a period of side length $L$.
Surprisingly, it is found that~\cite{GrimmElciZhouGaroniDeng2017}, in terms of $\bfU$, the plateau of $G(\bfU)$ disappears 
and the Gaussian decaying behavior $G(\bfU) \sim \|\bfU\|^{2-d}$ extends to a distance $\xiu \sim L^{d/4}$,
which we call the unwrapped correlation length. 
Apparently, $\xiu$ diverges faster than the linear size $L$,
giving a vivid and unconventional picture that, at criticality, 
the SAW wraps around the periodic boundaries for many times
and the winding number diverges as $L^{(d-4)/4}$. The divergence of the correlation length for the 5D Ising model with PBC was observed in Ref.~\cite{JonesYoung2005}.

At the upper critical dimensionality $\dc =4$, the Gaussian and the complete-graph sets of exponents coincide, 
i.e., $(2,1\!+\!d/2)=(d/2,3d/4)=(2,3)$.
For such a case, logarithmic corrections are generally expected.
It was proposed in Ref.~\cite{Lv2019Two} that the free energy density of the O$(n)$ model with $n=1,2,3$ can be written as 
   \begin{align}
          f(t,h) = &L^{-4}\tilde{f}_0(tL^{\yt},hL^{\yh}) +  \nonumber \\ &                  
        L^{-4}\tilde{f}_1(tL^{\yt}(\ln L)^{\ythat},hL^{\yh}(\ln L)^{\yhhat} ),
    \label{equ:freeenergy_4d}
    \end{align}
where $(\yt,\yh)=(2,3)$ and the logarithmic exponents $\ythat = \frac{4-n}{2n+16}$, $\yhhat=1/4$ \cite{kenna2013universal,kenna2004finite}. 
As in Eq.\eqref{equ:freeenergy_5d}, there simultaneously exist two terms in Eq.\eqref{equ:freeenergy_4d}; 
the former comes from the Gaussian fixed point
and describes the FSS of distance-dependent observables, and
the latter describes the ``background" contributions for the FSS of macroscopic quantities. However, $\tilde{f}_1$ can no longer be regarded as an exact counterpart of the FSS of complete graph, and thus, apart from the presence of multiplicative logarithmic corrections, this might lead to other effects, e.g., deviation of some universal quantities from their complete-graph counterparts.

In Ref.~\cite{Lv2019Two}, the authors simulated the O$(n)$ model with $n=1,2,3$ 
and measured the Fourier mode of susceptibility $\chi_{\bfk}$. 
They numerically found that $\chi_{\bfk \neq 0} \sim L^{2\yh-d}$ and $\chi_{\bfk=0} \sim L^{2\yh-d}(\ln L)^{2\yhhat}$. 
For the two-point function, the authors observed that
   \begin{equation}
    \label{equ:4dSAWfinitegx}
    G(\bfx) \sim 
    \begin{cases}
    \|\bfx\|^{-2} &\text{if}~\|\bfx\| \leq  O\left(\frac{L}{(\ln L)^{\yhhat}}\right)\\ 
    L^{-2}(\ln L)^{2\yhhat} &\text{if}~\|\bfx\| \geq  O\left(\frac{L}{(\ln L)^{\yhhat}}\right),
    \end{cases}
    \end{equation} 
and estimated $\yhhat = 0.25(5)$, in agreement with  the predicted value $1/4$.
Similarily to the $d>\dc$ case, one also expects that there exists an unwrapped correlation length $\xiu$, which diverges logarithmically faster than $L$ as $\xiu \sim L (\ln L)^{\yhhat}$. Consistent picture was observed for the correlation length of the 4D Ising model with PBC in Ref.~\cite{JonesYoung2005}.

In Ref.~\cite{Lv2019Two}, little attention was paid to the scaling behavior of thermal quantities. 
For example, the logarithmic divergence of the specific heat at the critical point, $C \sim (\ln L)^{2 \ythat}$ 
as predicted from Eq.~\eqref{equ:freeenergy_4d}, was not observed in Ref.~\cite{Lv2019Two}. 
In Ref.~\cite{lundow2009critical}, the authors simulated the 4D Ising model at the critical point 
and observed that the specific heat is still bounded with system sizes up to $L=80$, 
which is contradictory to the prediction from Eq.~\eqref{equ:freeenergy_4d}. 
Thus, more evidences are needed to clarify the critical scaling of the specific heat,
and to support Eq.~\eqref{equ:freeenergy_4d} consequently.
It would be also interesting to study the FSS of the unwrapped correlation length $\xiu$.

By universality arguments, we expect that Eq.~\eqref{equ:freeenergy_4d} can also be applied to SAW with $\ythat = 1/4$, 
which is larger than those $n>0$ cases. 
For SAW, larger system sizes are achievable in simulations than spin models, 
since one only needs to record a single walk instead of the whole lattice. 
Also, an irreversible Berretti-Sokal(BS) algorithm was developed in~\cite{hu2017irreversible}, 
which was shown to be particularly efficient for simulating SAWs on high-dimensional lattices. 
These advantages have been utilized in Refs.~\cite{GrimmElciZhouGaroniDeng2017,Grimm2018,ZhouGrimmFangDengGaroni2018} 
to confirm the form of $G(\bfx)$ in Eq.\eqref{equ:5DSAW_finite_gx} and $G(\bfU)$. 
Moreover, in the high-temperature phase ($z < \zc$) and in the thermodynamic limit, 
the specific heat equals zero since the energy density of SAW (walk length divided by the volume) is zero. 
Therefore, compared with spin models, the specific heat of SAW has no regular part at the critical point, 
and is expected to suffer less finite-size corrections.

In this paper, we simulated the SAW in 4D with PBC and the linear system size is up to $L = 768$. 
We found that, among various quantities, the unwrapped end-to-end distance (unwrapped correlation length) $\xiu$ 
suffers surprisingly weak finite-size corrections, from which we obtain a precise estimate of the critical point $\zc= 0.147\,622\,380(2)$, 
improving the existing result $0.147\,622\,3(1)$ \cite{owczarek2001scaling} by 50 times. 
At the estimated $\zc$, we observed the expected scaling for all the measured quantities as by Eq.~\eqref{equ:freeenergy_4d}, 
such as the mean walk length $N \sim L^{2}(\ln L)^{\ythat}$ and the susceptibility $\chi \sim L^{2}(\ln L)^{2\yhhat}$. 
In particular, we clearly observe the specific heat diverges as $C\sim (\ln L)^{2\ythat}$, which, together with the results in Ref.~\cite{Lv2019Two}, provides a complete numerical support to Eq.~\eqref{equ:freeenergy_4d}.
Our data analysis estimates that $\yhhat = 0.251(2)$ and $\ythat = 0.25(3)$, in agreement with the expected value $1/4$.

The $\widetilde{f}_1$ term in Eq.~\eqref{equ:freeenergy_4d} acts as the solely-$L$-dependent background contribution, 
which would vanish under Fourier transformation. 
So, the Fourier modes of the macroscopic quantities are expected to follow the prediction from the $\widetilde{f}_0$ term.
Indeed, we measure the Fourier modes of the mean walk length $N_{\bfk}$, the specific heat $C_{\bfk}$ and 
the susceptibility $\chi_{\bfk}$ with $\bfk\neq \origin$, and observe that 
$N_{\bfk} \sim L^2$, $\chi_{\bfk} \sim L^2$ and $C_{\bfk}$ tends to a constant, all consistent with the Gaussian fixed-point predictions.

The remainder of this article is organized as follows. 
In Sec.\ref{Observables and simulation}, we define the variable-length ensemble of SAW in 4D torus 
and introduce the algorithms, sampled observables and their expected finite-size scaling behavior. 
Numerical results are presented in Sec.\ref{Results}. We end with a discussion in Sec.\ref{discussion}.

  \section{Observables and FSS analysis}
  \label{Observables and simulation}

   \subsection{Model and algorithm}
   \label{Algorithm}

We use an irreversible version of the Berretti-Sokal algorithm~\cite{hu2017irreversible} to 
simulate self-avoiding walks on four-dimensional hypercubic lattices with PBC. 
We first introduce some notations. Let
\begin{equation}
\mathbb{T}_L = 
\begin{cases}
\left[\frac{1-L}{2}, \frac{L-1}{2} \right]^d\;, & \ \text{if}\ L \ \text{is odd} \\
\left[1 - \frac{L}{2}, \frac{L}{2} \right]^d\;, & \ \text{if}\ L \ \text{is even} \\
\end{cases}
\end{equation}
be the $d$-dimensional boxes with linear size $L$ and periodic boundary conditions. 
Let $\omega= (\bmomega_0, \bmomega_1, \bmomega_2, \cdots, \bmomega_{\ell})$ be an $\ell$-step self-avoiding walk, 
where each $\bmomega_i \in \mathbb{T}_L$ and $\|\bmomega_i - \bmomega_{i+1}\| = 1$. 
For simplicity, $\ell$ is used to specify the walk length $\ell=|\omega|$.
Let $\scrW$ be the set of all self-avoiding walks on $\mathbb{T}_L$ rooted at the origin (fixing $\bmomega_0=\origin$). 
We shall simulate with a fixed fugacity $z$, i.e., 
in the variable-length (grand canonical) ensemble of the self-avoiding walk model on $\mathbb{T}_L$, defined by choosing a SAW $\omega \in \scrW$ via the following probability,
\begin{equation}
\label{equ:SAW_4D}
\pi(\omega) = \chi^{-1} \cdot z^{\ell} \;, \ \  \forall \ \omega\in\scrW \;,
\end{equation}
where the susceptibility $\chi$, as defined in Eq.~(1), acts as a normalization factor.
On a high-dimensional hypercubic lattice with coordination number $\Delta=2d$, 
the critical fugacity is known to be $z_c \approx 1/(2d-1) \ll 1$.

The celebrated Metropolis method can be straightforwardly used to simulate
the SAW by Eq.~(\ref{equ:SAW_4D}). 
Among the nearest neighbors of the endpoint $\bmomega_\ell$, 
one of them is randomly chosen, say $\bfx$, and the endpoint is proposed to move there.
For $z<1$, if the chosen vertex $\bfx$ happens to be at $\bmomega_{\ell-1}$, 
one simply erases the SAW by one step ($\ell \leftarrow \ell-1$), 
and moves the endpoint there ($\bmomega_\ell \leftarrow \bfx$). 
Otherwise, if the self-avoiding condition is satisfied ($\bfx$ is not on the walk), 
then the proposal of $\bmomega_\ell \leftarrow \bfx$ is to 
increase the length by one step, $\ell \leftarrow \ell+1$, 
and the acceptance probability is $P=z$.
Near $z_c$, about $d$ trials are needed in order to have one successful update.
The BS algorithm~\cite{berretti1985new} introduced a simple trick to overcome this problem.
One first randomly chooses either the length-increasing ($+$) or 
the length-decreasing ($-$) operation with half-by-half probability.
For operation $(-)$, the endpoint $\bmomega_\ell$ is directly 
moved to the next-to-the-end point $\bmomega_{\ell-1}$, 
and the acceptance probability is $P^{(-)} = \textrm{min} \{ 1, 1/z(\Delta-1) \}$. 
For operation $(+)$, one uniformly at random chooses one of $(\Delta-1)$ neighbors 
(not including $\bmomega_{\ell-1}$) of $\bmomega_\ell$, and 
the acceptance probability is $P^{(+)} = \textrm{min} \{ 1, z(\Delta -1) \}$.
The detailed-balance condition can be easily proved. 
Near criticality $z_c \approx 1/(\Delta-1)$, 
the acceptance probabilities are close to 1, and thus
the BS algorithm gains a speeding-up factor about $d$, 
compared with the standard Metropolis algorithm.

The simulation efficiency can be further significantly improved by introducing an irreversible procedure,
in which the balance condition is satisfied but the detailed-balance condition is broken.
For the irreversible BS method, each SAW state in the configuration space $\scrW$ is duplicated 
into two copies, specified by an auxiliary state, $(+)$ and $(-)$, respectively. 
Within the extended configuration space $\widetilde{\scrW} = \scrW \otimes\{-,+\}$,
three types of update operations are introduced: the length-increasing, 
the length-decreasing, and the switching operation. 
The first two operations are the same as those in the BS algorithm, and 
the switching operation is to switch between the auxiliary states 
while the walk is kept unchanged. 
For any SAW with auxiliary state $(+)$, only the length-increasing and the switching operation are allowed; 
for any SAW with auxiliary state $(-)$, only the length-decreasing and the switching operation are allowed.

With the above definitions, the irreversible BS algorithm~\cite{hu2017irreversible} can be easily formulated as follows. 
For a SAW with $(-)$, the length-decreasing operation is performed with probability $P^{(-)}$, 
and, otherwise, the switching operation. 
Similarly, for a SAW with $(+)$, the length-increasing operation is carried out 
with probability $P^{(+)}$, and the switching operation, otherwise.
It can be derived from the balance condition that the acceptance probabilities, 
$P^{(+)}$ and $P^{(-)}$, are the same as those in the reversible BS algorithm. 
At the critical point, since $\zc (2d-1)$ is slightly larger than 1, 
one has $P^{(+)}=1$ and $P^{(-)}$ less but very close to 1;
for instance, $P^{(-)} \approx 0.97$ for the critical 4D SAW.
In this case, the simulation procedure is the following. 
The SAW keeps growing untilit hits itself, and then, the length-decreasing operation 
persists until a rejection event occurs with a small probability 
or the SAW is completely erased.
This leads to large probability flow circles, and the diffusive 
feature of random updates in the Metropolis and the BS algorithm is suppressed 
and partly replaced by ballistic-like behavior.
For the critical 4D SAW, in comparison with the Metropolis algorithm, 
the efficiency is improved by more than 100 times~\cite{hu2017irreversible}. 
For the complete graph, the irreversible BS algorithm 
is qualitatively more efficient than the reversible methods, 
as for the irreversible worm algorithm~\cite{LiftedWorm}.

Finally, it is mentioned that a hash-table technique is implemented in order to 
efficiently check the self-avoiding condition.

\subsection{Sampled quantities}
\label{Sample quantites}
In simulations, besides the conventional coordinates $\bfx$, the unwrapped coordinates of the SAW are dynamically calculated and stored in computer memory.
Initially (empty lattice), $\bfU = \origin$. At each step of updating the walk, add ${\bfe}_i$ ($-{\bfe}_i$) to $\bfU$ if the walk moves along (against) the $i$th direction.
Here $\bfe_i$ is the unit vector in the $i$th direction.

After thermalization, measurements of various observables are taken in every $L^2$ Monte Carlo steps. 
Given a self-avoiding walk $\omega$, we sample the following observables. 
For these Fourier modes, we sample with the wave vector $\bfk(k,i) = \frac{2\pi k}{L}{\bfe}_i$ where $i = 1,2,3,4$ and $k=1,2,3$. Our data show that Fourier modes using $\bfk(2,1)$ suffer slightly less finite-size corrections, so in the following discussion we fix the wave vector $\bfk = \bfk(2,1)$.
 \begin{enumerate} [label=(\alph*)]
   \item The walk length $\scrN = |\omega|$ 
         and its Fourier mode $\scrN_{\bfk} = \sum_{j=1}^{\scrN } e^{i\bfk\cdot\bmomega_j} $;
   \item The return-to-origin indicator $\scrD_0 = \ind(\scrN = 0)$ 
         and a Fourier mode $\scrF_{\bfk} = e^{i\bfk \cdot \bmomega_{\scrN}}$.
   \item The end-to-end distance $\scrR = \| \bmomega_{\scrN} \|$ 
         and its unwrapped version $\scrR_{\rm u}:=\sum_{i=1}^d |\bfU(\omega)_i|/d$, i.e., the averaged absolute value of the coordinates of $\bfU(\omega)$. 
  \end{enumerate}
The measurement occupies little computer time, since the measured observables are dynamically updated.

  We then obtain the ensemble average ($\langle \cdot \rangle$) of the following quantities:
  \begin{enumerate}[label=(\roman*)]
  \item The mean walk length $N = \langle \scrN \rangle$, an energy-like quantity, 
        and its variance $C = \frac{1}{V}(\langle \scrN^2 \rangle -\langle \scrN \rangle ^2)$, 
        which is the analog of the specific heat;
  \item The Fourier mode of walk length $N_{\bfk} = \langle | \scrN_{\bfk}| \rangle $ 
        and the specific heat $C_{\bfk} = \frac{1}{V}(\langle |\scrN_{\bfk}|^2 \rangle  - \langle |\scrN_{\bfk}| \rangle^2 ) $;
  \item The susceptibility $\chi = 1/\langle \scrD_0 \rangle$. 
        This equality holds since $\langle \scrD_0  \rangle = \frac{\sum_{\omega: |\omega | = 0} 
        z^{ |\omega |}} {\sum_{\omega} z^{ |\omega |}} = 1/\chi$. 
        The Fourier mode of susceptibility $\chi_{\bfk} = \langle | \scrF_{\bfk} | \rangle $/$\langle \scrD_0 \rangle$ ;
  \item The mean end-to-end distance $\xi = \langle \scrR \rangle $ 
        and the mean unwrapped end-to-end distance $\xi_{\rm u} = \langle \scrR_{\rm u} \rangle$;
  \item The Binder ratio $Q_N =\frac{\langle \scrN^2 \rangle }{\langle \scrN \rangle ^2}$.
  \end{enumerate}
   
   \subsection{Finite-size scaling analysis}
   \label{The FSS analysis}

By universality argument, we expect the scaling form of the finite-size free energy density $f(t,h)$  in Eq.~\eqref{equ:freeenergy_4d}, 
conjectured for the ${\rm O}(n)$ model with $n=1,2,3$, can be extended to the SAW model ($n=0$). 
When $h=0$, the FSS formulas for the susceptibility $\chi$, the mean walk length $N$ and its variance $C$ are
 	\begin{eqnarray}
 	\label{equ:chit}
	 \chi(t,L) &\sim & L^2 (\ln L)^{2\yhhat} \widetilde{\chi}_0[tL^{\yt}(\ln L)^{\ythat}] 
          +  L^2  \widetilde{\chi}_1\left(t L^{\yt}\right),	\nonumber \\
	 N(t,L)  &\sim & L^{2} (\ln L)^{\ythat} \widetilde{N}_0[tL^{\yt}(\ln L)^{\ythat}] 
          +L^{2}\widetilde{N}_1\left(t L^{\yt}\right),	\nonumber \\
	 C(t,L) &\sim &  (\ln L)^{2\ythat} \widetilde{C}_0[tL^{\yt}(\ln L)^{\ythat}] 
          +  \widetilde{C}_1\left( tL^{\yt} \right),	
 	\end{eqnarray}
 	where $t = z-z_{\rm c}$, $(\yt,\yh) = (2,3)$, $(\ythat,\yhhat)=(1/4,1/4)$ and $\widetilde{\chi}_0(\cdot)$, $\widetilde{\chi}_1(\cdot)$, $\widetilde{N}_0(\cdot)$, $\widetilde{N}_1(\cdot)$, $\widetilde{C}_0(\cdot)$, $\widetilde{C}_1(\cdot)$ are scaling functions.

We now derive the FSS of the unwrapped end-to-end distance $\xi_{\rm u}$. 
Consider an \emph{unwrapped} two-point function $\widetilde{G}({\bf u}) :=\sum_{\omega:\bfU(\omega)={\bf u}} {z}^{|\omega|} $,  
we expect that, 
for $d\geq d_c$, $\widetilde{G}({\bf u}) \sim \| {\bf u} \|^{2-d}g(\|{\bf u}\|/\xiu)$ 
where the function $g(x)$ decays sufficiently fast to zero as $x\rightarrow \infty$ 
such that the integral $\int_{0}^\infty x^\alpha g(x) <\infty$ for all $\alpha \in (0, \infty)$. 
It then follows that
\begin{eqnarray}
\chi = \sum_{{\bf u}} \widetilde{G}({\bf u}) \approx \int_0^\infty t g(t/\xi_{\rm u}) \d t \sim \xi^2_{\rm u}
\end{eqnarray}
Since $\chi\sim L^2(\ln L)^{2\yhhat}$, it follows that $\xi_{\rm u} \sim L(\ln L)^{\yhhat}$. Near the critical point, we expect that
\begin{equation}
\label{equ:Rut}
\xi_{\rm u}(t,L) \sim L(\ln L)^{\yhhat} \widetilde{\xi}_{\rm u}[tL^{y_{\rm t}}(\ln L)^{\ythat}]
\end{equation}
where $\widetilde{\xi}_{\rm u}(\cdot)$ is the scaling function.

The FSS of Fourier-mode quantities are expected to follow the Gaussian fixed-point predictions, that is,
\begin{equation}
    \label{equ:fourier_transform}
       \chi_{\bfk} \sim L^2 \;, \ \   N_{\bfk} \sim L^2 \;, \ \   C_{\bfk} \sim \text{const}.
\end{equation}

\begin{table*}
\centering
\begin{tabular}{|c|c|c|c|c|c|c|c|c|c|}
\hline 
$L_{\rm min}$ &$\yt$ 	&$\ythat$ 	&$z_c$ 	&$c_0$ 	&$c'_0$ &$a_0$ 	&$a_1$ 	&$a_2$ 	 & $\text{chi}^2/{\rm DF}$ \\ 
\hline
48  &2.005(5)   &1/4    &0.147\,622\,379\,6(8) & 0 &0 	&  0.443\,52(4)   	&0.83(2)	&1.11(6)  &16.2/17\\ 
64  &2.005(5)   &1/4    &0.147\,622\,379\,7(8)& 0 &0 	&  0.443\,53(5)   	&0.83(2)   	&1.11(6)  &16.0/15\\ 
96  &2.005(6)   &1/4    &0.147\,622\,380\,0(8)& 0 &0 	&  0.443\,62(7)   	&0.83(2)  	&1.11(6)  &10.8/11\\ 
\hline 
48  &2  &0.28(3)   &0.147\,622\,379\,6(8) & 0 &0 &  0.443\,52(4)       &0.81(4)   	&1.07(9)   &16.1/17\\ 
64  &2  &0.28(3)   &0.147\,622\,379\,7(8) & 0 &0 &  0.443\,53(5)       &0.81(4)   	&1.07(9)   &15.9/15\\ 
96  &2  &0.28(3)   &0.147\,622\,380\,0(8) & 0 &0 &  0.443\,62(7)       &0.81(4)   	&1.1(1)    &10.7/11\\ 	
\hline	
48  &2  &1/4  &0.147\,622\,379\,7(8)  & 0 &0 &0.443\,52(4) &0.850(2)     &1.16(2)   & 17.1/18 \\ 	     
64  &2  &1/4  &0.147\,622\,379\,7(8)  & 0 &0 &  0.443\,53(5)  &0.850(2)   	&1.16(2)   &16.8/16\\ 
96  &2  & 1/4 &0.147\,622\,380\,0(8)& 0 &0 &0.443\,63(7) &0.850(2)     &  1.16(2)       &11.5/12\\ 	
\hline
48  &2  &1/4  &0.147\,622\,380\,6(9)  & $-0.04(2)$ &0 &0.444\,6(5)  &0.852(2)     &1.15(3)   & 12.8/17 \\ 	     
64  &2  &1/4  &0.147\,622\,381(1)       & $-0.08(3)$ &0 &0.445\,6(8)  &0.853(3)     &1.14(4)   &9.6/15\\ 
\hline
48  &2  &1/4  &0.147\,622\,379\,6(8)  & 0 & $-0.5(5)$ &0.443\,52(4)  &0.87(2)     &1.24(8)   & 16.1/17 \\ 	     
64  &2  &1/4  &0.147\,622\,379\,7(8)  & 0 & $-0.5(5)$ &0.443\,53(5)  &0.87(2)     &1.24(8)   &15.9/15\\ 
\hline
\end{tabular}
\caption{Fitting results for the unwrapped end-to-end distance $\xiu$, using the ansatz Eq.~\eqref{equ:fits_zc} with $m=2$, and $b_1,b_2, c_1 = 0$.}
\label{tab:fit_zc_Ru} 
\end{table*}

\section{Results}
\label{Results}

In this section, we provide numerical results to support the scaling behavior of various quantities conjectured in Sec.~\ref{The FSS analysis}. We perform least-square fits of our Monte Carlo data to the expected ansatz. As a precaution against correction-to-scaling terms that we miss including in the fitting ansatz, we impose a lower cutoff $L \geq L_{\rm min}$ on the data points admitted in the fits, and systematically study the effect on the residuals (denoted by $\text{chi}^2$) by increasing $L_{\rm min}$. In general, the preferred fit for any given ansatz corresponds to the smallest $L_{\rm min}$ for which the goodness of the fit is reasonable and for which subsequent increases in $L_{\rm min}$ do not cause the $\text{chi}^2$ value to drop by vastly more than one unit per degree of freedom(DF). In practice, by ``reasonable" we mean that $\text{chi}^2/{\rm DF} \lesssim 1$. The systematic error is obtained by comparing estimates from various reasonable fitting ansatz.

  \begin{figure}[t]
   	\centering
   		\includegraphics[scale=0.65]{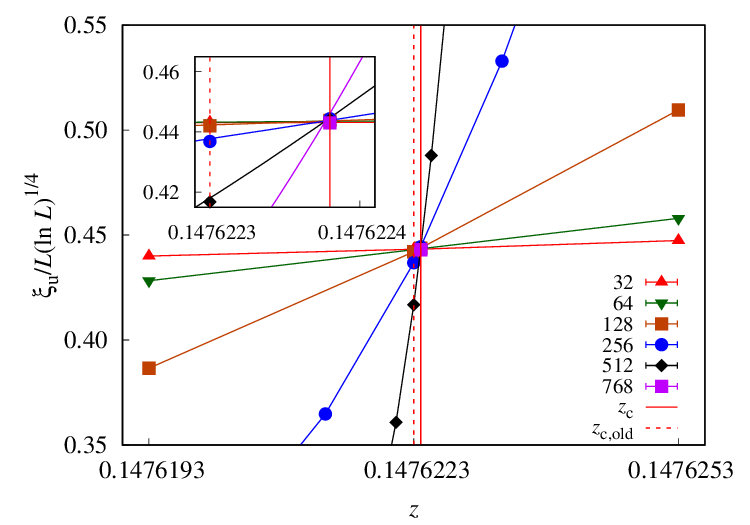}
   	\caption{Plot of the unwrapped end-to-end distance $\xi_{\rm u}$, rescaled by $L (\ln L)^{1/4}$, near the critical point and for various system sizes. The red vertical line indicates the new estimate of $\zc$, while the dashed vertical line is the central value of the previously best estimate of $\zc = 0.147\,622\,3$. The inset zooms in the intersection region, to clearly show their deviation.}
   	   	\label{figRut}
   \end{figure}

We first carry out extensive simulations at the previously best estimate $\zc = 0.147\,622\,3$~\cite{owczarek2001scaling} with system sizes up to 512. However, from these data, the estimates of $\ythat$ (from $N$ and $C$) and $\yhhat$ (from $\chi$) are inconsistent with their expected value $1/4$. The reason might be the precision of $\zc$ is not high enough. So we next carry out simulations around $\zc$ to estimate $\zc$ to a higher precision. 

\subsection{Estimate of the critical point $z_c$}
\label{Estimate of critical point}

As suggested by the scaling behavior of $\xi_{\rm u}$ in Eq.~\eqref{equ:Rut}, if one plots $\xi_{\rm u}/[L (\ln L)^{\yhhat}]$ versus $z$ for various system sizes, then at the critical point an intersection is expected for large $L$. Indeed, as shown in Fig.~\ref{figRut}, by letting $\yhhat=1/4$ one can see an excellent intersection point for $L\geq 32$. This suggests that $\xi_{\rm u}$ suffers very weak finite-size corrections and thus is a nice quantity for estimating $\zc$. In the inset of Fig.~\ref{figRut}, we can see that the intersection point is around $0.147\,622\,380$, which is clearly larger than the central value of the previously best estimate, as indicated by the red dashed line.

To estimate $\zc$ systematically, we perform the least-square fits of the $\xi_{\rm u}$ data onto the following ansatz,
\begin{align}
\label{equ:fits_zc}
\nonumber
\frac{\xi_{\rm u}}{L \left(\ln L + c_0 \right)^{1/4}}  & = \sum_{k=0}^{m} a_k (z - \zc)^k (L^{\yt} (\ln L + c'_0)^{\ythat})^k \\
&+  b_1 L^{y_1}  + b_2  L^{y_2} \nonumber \\
& + c_1 (z - \zc) L^{\yt+y_1}(\ln L + c'_0)^{\ythat} \;.
\end{align}
Here $m$ is the highest order we keep in the fitting ansatz, from the Taylor expansion of $\widetilde{\xi}_{\rm u}$ around $z=\zc$. The constants $c_0, c'_0$ are nonuniversal constants which are commonly used when fitting the data at the upper critical
dimension where logarithmic corrections are involved~\cite{gruneberg2004universal}. The terms $b_1,b_2$ account for the finite-size corrections with $y_2 < y_1 < 0$. The $c_1$ term accounts for the crossing effect between finite-size corrections and the scaling variable $(z-z_c) L^{\yt} (\ln L)^{\ythat}$.

We start with fixing $c_0, c'_0 = 0$.
From our fits, we notice that leaving both $\yt$ and $\ythat$ free cannot produce stable results. So we start with fitting by fixing $\yt$ to its expected value $2$ and leaving $\ythat$ free. We first try the fits with $m=2$ and without including any correction terms in the ansatz, that is, setting $b_1,b_2,c_1$ to zero. The result shows that ${\rm chi}^2/{\rm DF} < 1$ when $L_{\rm min}=48$, so indeed $\xi_{\rm u}$ suffers quite weak finite-size corrections. This fit gives that $\zc = 0.147\,622\,379\,8(10)$ and $\ythat = 0.28(3)$, consistent with the expected value $1/4$. The coefficients $a_k$ are found to be consistent with zero when $k\geq 3$. Thus, in the following, we fix $m=2$. We then try to fit by including one correction term $b_1L^{y_1}$. Leaving $y_1$ free cannot produce stable fits, which is expected since the corrections are very weak ($y_1$ is small). Fixing $y_1$ to either $-1$ or $-2$ in the ansatz produce  consistent estimate  $\zc = 0.147\,622\,380\,6(12)$  and $\zc = 0.147\,622\,380\,0(9)$ respectively. Consistent estimates are also obtained when fitting with fixing $y_1 = -1$ and $y_2 = -2$. In all scenarios above, including the crossing-effect term to the ansatz shows that $c_1$ consistent with zero and its effect to the estimates of other parameters is negligible. 

We then perform the fits with $\ythat$ fixed at $1/4$ and $\yt$ free, and follow the similar procedure described above.  In this scenario, our fits again produce consistent estimate for $\zc$ and $\yt$. Similar results are also obtained from the fits with fixing both $\yt = 2$ and $\ythat = 1/4$. We show some fitting details in Table~\ref{tab:fit_zc_Ru}.

\begin{figure}[t]
   	\centering
   		\includegraphics[scale=1]{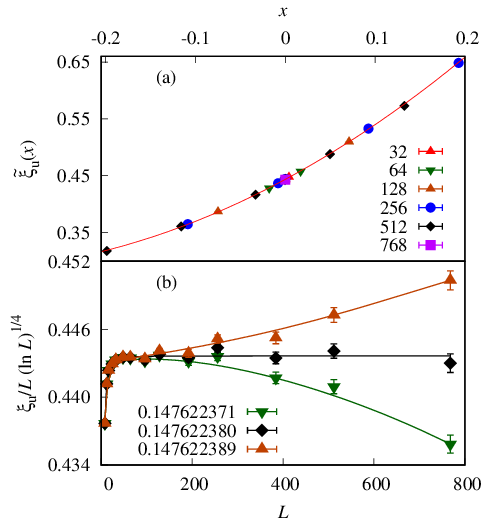}
   	\caption{(a) Plot of the scaling function $\widetilde{\xi}_{\rm u}(x)$, where $x = (z-\zc)L^{2}(\ln L)^{1/4}$. The red curve corresponds to our preferred fit of the $\xiu$ data to the ansatz Eq.~\eqref{equ:fits_zc}. (b) Plot of $\xiu/[ L(\ln L)^{1/4}]$ versus $L$ for fixed values of $z$. The curves correspond to our preferred fit of the data by the ansatz Eq.~\eqref{equ:fits_zc}. We note that, the central value of the previously best estimate of $\zc$ deviates away from our estimate about 40 times of our quoted error bar.}
   	   	\label{fig:demon_zc}
   \end{figure}

We then try the fits with leaving $c_0, c'_0$ free. However, leaving both $c_0, c'_0$ free in the fits cannot produce stable results. We next try to fix $c'_0 = 0$ and leave $c_0$ free. Reasonable fits are obtained at $L_{\rm min} = 48$, which gives that $\zc = 0.147\,622\,380\,6(9)$ and $c_0 = -0.04(2)$. The estimate of $\zc$ is consistent with the fixing $c_0, c'_0 = 0$ case. Finally, we try the fits with fixing $c_0 =0$ but leaving $c'_0$ free. Stable fits are obtained at $L_{\rm min} = 48$, which give $\zc = 0.147\,622\,379\,6(8)$ and $c'_0 = -0.5(5)$. Fitting details are shown in Table~\ref{tab:fit_zc_Ru}. Clearly, our fits suggest both $c_0$ and $c'_0$ are consistent with zero within two standard deviations, and thus the effect of including characteristic length to Eq.~\eqref{equ:fits_zc} is negligible. This means that $\xi_{\rm u}$ suffers very weak finite-size corrections, both in additive ($b_1, b_2 \approx 0$) and multiplicative logarithmic terms ($c_0, c'_0 \approx 0$). By comparing estimates from various ansatz, we conclude that $\zc = 0.147\,622\,380(2)$ and $\ythat = 0.27(4)$. In Fig.~\ref{fig:demon_zc}(a), we show the scaling function by plotting $\tilde{\xi}_{\rm u}(x)$ versus $x := (z-z_c)L^2(\ln L)^{1/4}$. As expected, the data collapse nicely onto the curve which corresponds to our preferred fitting to the ansatz Eq.~\eqref{equ:fits_zc}.

To demonstrate the validity of our estimate of $\zc$, we plot $\xiu/[ L(\ln L)^{1/4} ]$ versus $L$ for different values of $z$, shown in Fig.~\ref{fig:demon_zc}(b). As expected, the data with $z=0.147\,622\,380$ tend to a horizontal line, while the data with $z$ about five standard deviations away from our estimate clearly bend up and down, respectively. The convergence towards a horizontal line at $\zc$ also suggests that $c_0\approx 0$, consistent with our fitting results.

We also perform least-square fits to the data of other quantities, such as $Q_N$, $\chi$, $C$ and $N$, to estimate $\zc$.  Our results show that, compared with $\xi_{\rm u}$, these quantities suffer stronger finite-size corrections and produce consistent estimates of $\zc$ but with larger error bars. Fitting details for these quantities are omitted here.
   
  \subsection{Scaling behavior at $z_{\rm c}$}  
   \label{scaling behavior at critical point}
   
      \begin{figure}[t]
	\centering
	\includegraphics[scale=1.0]{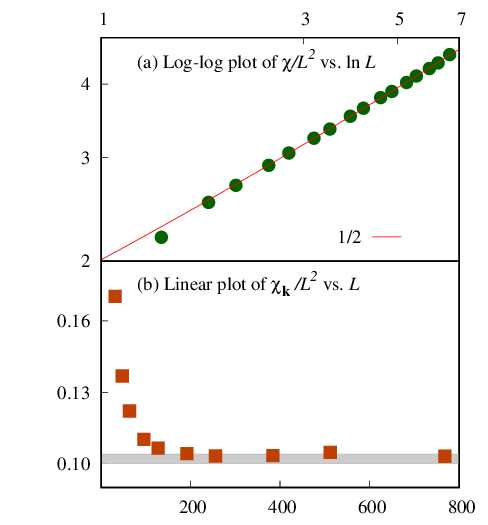}
	\caption{The scaling behavior of susceptibility $\chi$ (a) and its Fourier mode $\chi_{\bfk}$ (b), at the critical point. The line in the top figure has slope $1/2$, and the shadow in the bottom figure indicates our estimate of the value that $\chi_{\bfk}/L^2$ converges to. All data points here have error bars which are smaller than the size of data points.}
	\label{fig:chit}
\end{figure}

We then study the scaling behavior of various quantities at $\zc$, to estimate the logarithmic correction exponents $\yhhat, \ythat$ and the universal Binder cumulant $Q_N$. We first study $\xi_{\rm u}$, which is expected to scale as $L (\ln L)^{\yhhat}$ at $\zc$. To estimate $\yhhat$, we perform the least-square fits of the $\xi_{\rm u}/L$ data to the ansatz
\begin{equation}
\label{Eq:fitansatz_xiu_zc}
\scrO = (\ln L)^{\hat{y}_{\scrO}}(a_0 +b_1L^{y_1}) \;.
\end{equation}
Here the exponent $\hat{y}_{\scrO} = \yhhat$. Again, without including any correction terms (fixing $b_1= 0$), we obtain stable fits when $L_{\rm min} = 48$, and it gives $\yhhat = 0.252(2)$, in excellent agreement with the expected value $1/4$. Including only one correction term $b_1L^{y_1}$ in the fitting ansatz with $y_1$ free produces stable fits even when $L_{\rm min} = 6$, which gives $\yhhat = 0.2518(8)$ and $y_1 = -2.5(1)$. We also try to fit by fixing $y_1 = -5/2$ and consistent estimate of $\yhhat$ is obtained. By comparing estimates from various ansatz, we have $\yhhat = 0.251(2)$. The details of our fits are shown in Table~\ref{tab:fit_Ru}.

  \begin{table}[H]
   	\centering
   	\begin{tabular}{|l|l|l|l|l|l|}
   		\hline
   		$L_{\rm min}$	&$\yhhat$	&$a_0$	&$b_1$	&$y_1$	&$\text{chi}^2$/DF \\
   		\hline
   		6	&0.2518(8)	& 0.4424(5)	&-1.1(1)	&-2.5(1)	&8.9/11 \\
   		8	&0.251(1)	 & 0.4426(7)	&-0.9(4)	&-2.4(2)	&8.7/10 \\
   		\hline
   		12  &0.2515(6)	&0.4426(5)	&-1.0(1)	&-5/2			&8.6/10 \\
   		16  &0.252(1)	&0.4423(6)	&-0.9(2)	&-5/2			&8.3/9 \\
   		\hline 	
   		48  &0.252(2)  &0.4425(9)  &0          & -          &7.2/7  \\
   		64  &0.252(2)   &0.442(1)   &0          & -          &6.9/6  \\
   		\hline 
   	\end{tabular}
   	\caption{The fitting results of the unwrapped distance $\xiu$ at the critical point.}
   	\label{tab:fit_Ru}
   \end{table}

We then study the susceptibility $\chi$ and its Fourier mode $\chi_{\bfk}$. From Eqs.~\eqref{equ:chit} and~\eqref{equ:fourier_transform}, we expect that $\chi \sim L^2(\ln L)^{2\yhhat}$ and $\chi_{\bfk} \sim L^2$, at $\zc$. Strong numerical evidence can be seen in Fig.~\ref{fig:chit}. In the top figure, we plot $\chi/L^2$ versus $\ln L$ in log-log scale, and clearly our data collapse onto a straight line with slope $1/2$. The bottom figure shows $\chi_{\bfk}/L^2$ converges as $L$ increases. To estimate $\yhhat$, we then perform least-square fits of the $\chi/L^2$ data to the ansatz
	\begin{equation}
    	\label{equ:fittingansatz}
    \scrO = a_0\left( \ln L + c_0 \right)^{\hat{y}_\scrO} + a_1.
    	\end{equation}
Here $\hat{y}_\scrO = 2\yhhat$. Leaving both $c_0$ and $a_1$ as free parameters cannot produce stable fits. Fixing one of them to zero and leaving the other one free produce consistent estimate of $\yhhat$. This is expected since, as subdominate terms, $c_0$ and $a_1$ have almost the same effect. Our fits give that $\yhhat = 0.24(2)$, consistent with the expected value $1/4$. The fitting detail is shown in Table~\ref{tab:fit_chiNC}. To check $\chi_{\bfk}/L^2$ converges to a nonzero constant, we perform a simple fit of the $\chi_{\bfk}/L^2$ data to Eq.~\eqref{Eq:fitansatz_xiu_zc} with $\hat{y}_{\scrO} = 0$, and it gives that $a_0 = 0.102(2)$.

\begin{figure}[H]
	\centering
	\includegraphics[width=1.0\columnwidth]{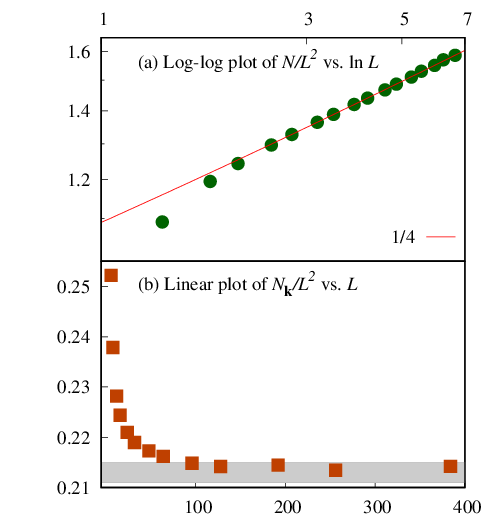}
	\caption{The scaling behavior of the mean walk length $N$(a) and its Fourier mode $N_{\bfk}$(b) at the critical point $z_c$. The line in the top figure has slope $1/4$, and the shadow in the bottom figure indicates our estimate of the value that $N_{\bfk}/L^2$ converges to. All data points here have error bars which are smaller than the size of data points.}
	\label{fig:ent}
\end{figure}

We then study the mean walk length $N$ and its Fourier mode $N_{\bfk}$, the expected scaling of which at $\zc$ are respectively $N \sim L^2 (\ln L)^{\ythat}$ and $N_{\bfk} \sim L^2$. Numerical evidence supporting their scaling is shown in Fig.~\ref{fig:ent}. In the top figure, we log-log plot $N/L^2$ versus $\ln L$ and our data collapse onto a line with slope $1/4$; the bottom figure shows that $N_{\bfk}/L^2$ converges as $L$ increases. Fitting $N/L^2$ and $N_{\bfk}/L^2$ using the same procedure as the susceptibility case, we obtain $\ythat = 0.26(2)$ and $N_{\bfk}/L^2$ converges to a constant $0.213(2)$.

\begin{figure}[H]
   	\centering
   	\includegraphics[width=1.0\columnwidth]{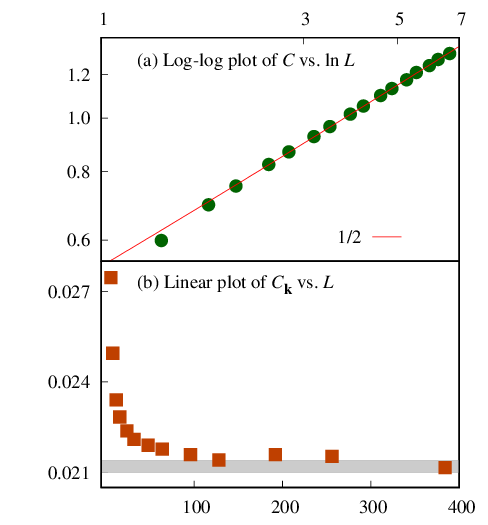}
   	\caption{The scaling behavior of the specific heat $C$(a) and its Fourier mode $C_{\bfk}$(b) at the critical point $z_c$. The line in the top figure has slope $1/2$, and the shadow in the bottom figure indicates our estimate of the value that $C_{\bfk}$ converges to. All data points here have error bars which are smaller than the size of data points.}
   	\label{fig:C2}
   \end{figure}

We then discuss the specific heat $C$ and its Fourier mode $C_{\bfk}$ at $\zc$. As shown in Eq.~\eqref{equ:chit}, we expect that $C$ diverges logarithmically as $C \sim (\ln L)^{2\ythat}$ for the ${\rm O}(n)$ model with $0\leq n\leq 3$. However, it is difficult to be numerically observed in the $n=1,2,3$ cases. For the self-avoiding walk, thanks to the precisely determined critical point and much larger achievable system sizes, we clearly observe the logarithmic divergence of the specific heat. As shown in Fig.~\ref{fig:C2}, the plot of $C$ versus $\ln L$ in log-log scale clearly shows that the data collapse onto a line with slope $1/2$. Perform least-square fits of the $C$ data to Eq.~\eqref{equ:fittingansatz} gives that $\ythat = 0.24(2)$, consistent with the expected value $1/4$. For the Fourier mode, our data shows that $C_{\bfk}$ converges to a constant $0.0212(5)$.
  
\begin{table}[H]
	\centering
	\begin{tabular}{|c|c|c|c|c|c|c|}
		\hline
		$\scrO$ & $L_{\rm min}$	&  &$c_0$	&$a_0$	&$a_1$	&$\text{chi}^2$/DF \\
		\hline
		\multirow{4}{*}{$\chi$}		
		&12	 &0.24(2)    & 0	&1.7(2)   &0.3(2)  	&6.5/10 \\
		&16   &0.25(2)    & 0   &1.5(2)   &0.5(2)   &5.1/9 \\
		\cline{2-7}
		&16	  &0.24(1)   & 0.3(2)	&1.79(9)   &0  	&4.9/9 \\
		&24    &0.23(2)    & 0.1(3)   &1.9(1)   & 0   &4.3/8 \\
		\hline
		\multirow{4}{*}{$N$}
		&16	&0.26(2)	& 0   &0.80(9)   &0.3(1)  	&10.1/9 \\
		&24  &0.29(4)    & 0   &0.7(1)    &0.4(1)      &8.7/8 \\
		\cline{2-7}
		&32	&0.24(1)	& 0.7(3)     &0.99(4)    &0    	&6.8/7 \\
		&48  &0.26(3)    & 1.2(7)      &0.94(8)    &0      &6.1/6 \\
		\hline
		\multirow{4}{*}{$C$}
		&24	&0.24(2)   & 0	   &0.53(6)   &0.01(6)  	&8.9/8 \\
		&32  &0.25(2)   & 0    &0.48(7)   &0.06(9)     &8.3/7 \\
		\cline{2-7}
		&24	&0.24(1)  & 0.0(1)	 &0.53(2)   &0  	 &8.9/8 \\
		&32  &0.25(1)  &  0.1(2)   &0.52(3)   &0     &8.2/7 \\
		\hline							
	\end{tabular}
	\caption{The fitting results of the susceptibility $\chi$, walk length $N$ and specific heat $C$ at the critical point. The third column is the estimate of $\yhhat$ for $\chi$, and $\ythat$ for $N$ and $C$.}
	\label{tab:fit_chiNC}
\end{table}

\subsection{Binder cumulant $Q_N$}

We finally discuss the Binder cumulant $Q_N$. For SAW on the complete graph with $V$ vertices, it was proved~\cite{deng2019length} that the mean walk length $N =\sqrt{2V/\pi} + o(1)$ and its variance is ${\rm var}\scrN = (1-2/\pi)V + o(1)$. It then follows that, on the complete graph,
\begin{equation}
Q_N = \frac{\langle \scrN^2 \rangle}{\langle \scrN \rangle^2}  = \frac{{\rm var}\scrN }{\langle \scrN \rangle^2} + 1 = \frac{\pi}{2} + o(1).
\end{equation}

However, for SAW at 4D, the situation is different. As we conjecture in Eq.~\eqref{equ:chit} and~\eqref{equ:Rut}, and numerically confirmed in Sec.~\ref{scaling behavior at critical point}, the mean walk length $N\sim L^2 (\ln L)^{\ythat}$ and $\xiu \sim L(\ln L)^{\ythat}$. Thus, the relation $N \sim \xiu^2$ fails to hold, and thus SAW at 4D behaves differently as the simple random walk. So, due to logarithmic corrections to scaling, the $\tilde{f}_1$ term in Eq.~\eqref{equ:freeenergy_4d} is not exactly the counterpart of the FSS on the complete-graph, and thus $Q_N$ may not take its complete-graph value.

   \begin{figure}[H]
 	\centering
 	\includegraphics[scale=0.67]{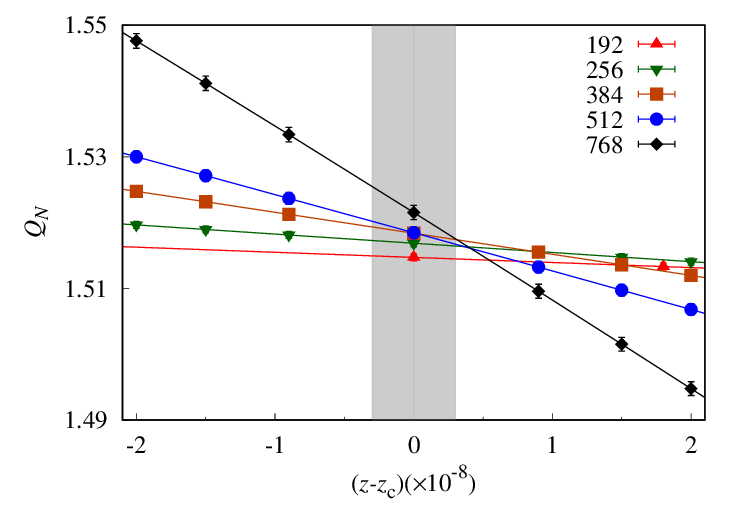}
 	\caption{Plot of the Binder cumulant $Q_N$ near the critical point, for various system sizes. The shadow column shows our estimate of $\zc$.}
 	\label{fig:qn_near_zc}
 \end{figure}

  \begin{figure}[H]
 	\centering
 	\includegraphics[scale=0.67]{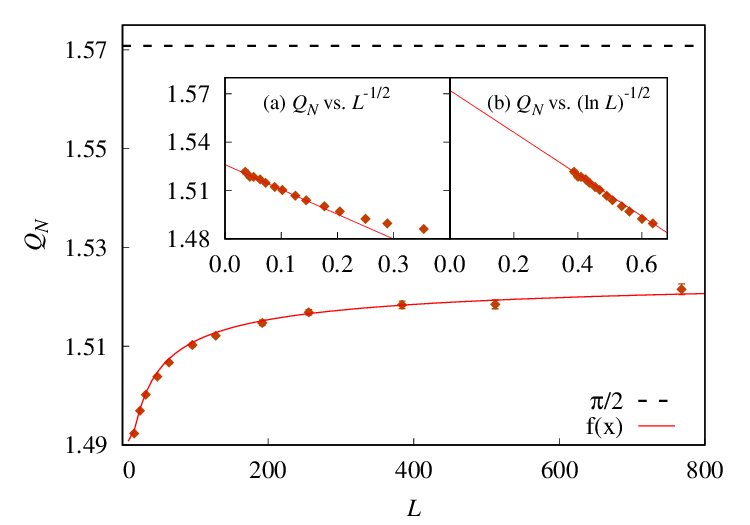}
 	\caption{Plot of the Binder cumulant $Q_N$ versus $L$ at the critical point. The curve corresponds to our preferred fit of the $Q_N$ data to the ansatz $Q_N = Q_0 + b_1 L^{y_1}$.	 The left and right insets plot $Q_N$ versus $L^{-1/2}$ and $(\ln L)^{-1/2}$, respectively. The two red lines are to guide the eyes.}
 	\label{fig:qn_at_zc}
 \end{figure}

We then examine the $Q_N$ data. In Fig.~\ref{fig:qn_near_zc}, we plot the $Q_N$ data near the critical point for various system sizes. Our estimate of $\zc$ is shown as the shadow column. As one can see, as $L$ increases, it is almost impossible to have an intersection located within our estimate of $\zc$ and giving the value $\pi/2$ in the vertical axis. This means that the value of $Q_N$ at $\zc$ may not take the complete-graph value $\pi/2$.

We next study the $Q_N$ data at the critical point. We first plot $Q_N$ versus $L$ in Fig.~\ref{fig:qn_at_zc}, which shows that, as $L$ increases, $Q_N$ likely converges to some constant smaller than $\pi/2$. To precisely estimate the value $Q_N$ converges to, we perform careful fits to the $Q_N$ data. We first try to fit without logarithmic corrections, using the ansatz

 \begin{equation}
 	   Q_N = Q_0 + b_1 L^{y_1} + b_2 L^{y_2}.
 	   \label{eq:QN_ansatz}
 \end{equation}
The data can be well fitted by including only one correction term (fix $b_2 = 0$), which gives $Q_0 = 1.528(3)$ and $y_1 = -0.43(8)$ when $L_{\rm min} = 48$. In the left inset of Fig.~\ref{fig:qn_at_zc}, we plot $Q_N$ versus $L^{-1/2}$ and the data collapse to a straight line, which also indicates the leading correction term of $Q_N$ is $L^{-1/2}$.  Consistent estimate of $Q_0$ is obtained when one more correction term $b_2 L^{-2}$ is added. Thus, without logarithmic corrections, our data analysis support that $Q_0$ is smaller than $\pi/2$.

We then try to fit with logarithmic corrections. As the right inset of Fig.~\ref{fig:qn_at_zc} shows, if we plot $Q_N$ versus $(\ln L)^{-1/2}$, then the data also collapse to a straight line which ends around $\pi/2$ as $L\rightarrow \infty$. This means the $Q_N$ data can also be fitted to the ansatz $Q_N = Q_0 + b_1 (\ln L)^{-1/2}$. Indeed, when $L_{\rm min} = 96$, our fits show that $Q_0 = 1.575(4)$. We also try to fit $Q_N$ to the ansatz $Q_N = Q_0 + b_1 (\ln L + c)^{-1/2}$. Stable fits are obtained when $L_{\rm min}=48$, and we have $Q_0 = 1.56(1)$. So, our data suggest that the estimate of $Q_0$ is consistent with $\pi/2$ when logarithmic corrections are taken into account.\par

In short, it is suggested that likely $Q_N$ at criticality deviates from the complete-graph value, although this can not be completely excluded due to the possible existence of logarithmic corrections. In previous literatures like Ref.~\cite{luijten1997interaction}, the Binder ratio and particularly its complete-graph value of the Binder ratio were used to estimate the critical point of the four-dimensional Ising model. Our data in Figs.~\ref{fig:qn_near_zc} and \ref{fig:qn_at_zc} imply that this might induce systematic deviations.
Nevertheless, whether $Q_N$ takes the complete-graph value $\pi/2$ at 4D  or less than $\pi/2$ remains an open question and needs further theoretical and numerical study. For completeness, we study in the Appendix the Binder cumulant $Q_N$ for $d=1,2,3,5$.
 \par

 \section{Discussion}
  \label{discussion}
  
In this paper, we studied the finite-size scaling of the self-avoiding walk model on four-dimensional hypercubic lattices with periodic boundary conditions. We first precisely locate the critical point $\zc$ by performing finite-size analysis to the unwrapped end-to-end distance $\xiu$. At the estimated $\zc$, we determine the logarithmic scaling behavior of various quantities: the susceptibility $\chi \sim L^2(\ln L)^{2\yhhat}$, the mean walk length $N \sim L^2 (\ln L)^{\ythat}$ and the specific heat $C \sim (\ln L)^{2\ythat}$. We estimate $\ythat = 0.25(3)$ and $\yhhat = 0.251(2)$, both in agreement with the expected value $1/4$. Moreover, for the Fourier modes with $\bfk \neq \origin$, we numerically observed that $\chi_{\bfk} \sim L^2$, $N_{\bfk} \sim L^2$ and $C_{\bfk}$ converges to a constant, following Gaussian fixed-point predictions. Our results provide strong numerical evidence to the conjectured finite-size scaling form of the free energy, shown in Eq.~\eqref{equ:freeenergy_4d}, for the ${\rm O}(n)$ universality class at 4D with $0\leq n \leq 3$.

We note that, compared with the previous understanding of FSS at $\dc$~\cite{kenna2004finite}, the significance of Eqs.~\eqref{equ:freeenergy_4d} and \eqref{equ:4dSAWfinitegx} can be seen from the following aspects. First, these two equations arise from a simple and beautiful picture i.e., the simultaneous existence of Gaussian fixed-point asymptotic and modified complete-graph asymptotic at $\dc$. Note that, the Gaussian fixed-point term is not merely a correction term, since it dominates the scaling behavior of Fourier-transformed quantities (with nonzero modes), determines the distance-dependent behavior of the two-point function, and also the quantities with scale much smaller than the system size. Therefore, without the Gaussian fixed-point term, the FSS ansatz of the free energy and even the entire physical picture at $\dc$ is incomplete. Moreover, Eq.~\eqref{equ:freeenergy_4d} can systematically predict the FSS of various thermal and magnetic quantities, which can explain all the existing numerical results in the literature.

From the finite-size analysis to the SAW data at 4D, we notice that whether the logarithmic scaling can be clearly observed depends sensitively on the precision the estimated critical point. In Ref.~\cite{Lv2019Two}, the logarithmic divergence of the specific heat for the $n$-vector model with $n=1,2,3$ was not clearly observed. The reason might be that the critical points for these cases have not been estimated to a high enough precision. For SAW, the precisely determined critical point attributes to a geometric quantity: the unwrapped end-to-end distance $\xiu$, which suffers extremely weak finite-size corrections. We note that geometric (random walk) representations are also available for these $n\geq 1$ cases~\cite{FernandezFrohlichSokal13}. So it is possible to find a quantity analogous to $\xiu$ which can be used to precisely estimate the critical points of these cases. Take the Ising model for example, one candidate of such a quantity is the unwrapped end-to-end distance of the Aizenmann random walk, see Ref.~\cite{DengGaroniGrimmZhou21} for the explicit definition. It would be interesting to perform a systematic finite-size analysis to the $n$-vector model under the geometric representation in four dimensions.

  \section{Acknowledgments}
  This work was supported by the National Natural Science Foundation of China (under Grant No.~11625522),
  the Science and Technology Committee of Shanghai (under grant No. 20DZ2210100),
  the National Key R\&D Program of China (under Grant No.~2018YFA0306501). We thank Hao Hu for valuable discussion. 
  
  \appendix
  \section{Binder cumulant $Q_{N}$ for other dimensions}
  \label{Appendix}
  
  	In this section, we discuss the dependence of  the critical Binder cumulant $Q_N$ on dimensionality $d$.
  	We first  simulate the SAW on the 2D, 3D and 5D hypercubic lattices  with PBC  at the critical points shown in Table~\ref{Tab:Num_saw}. The largest system sizes are up to $L=512, 192, 96$ and more than $10^7$, $10^6$, and $10^6$ samples for each system size are generated for 2D, 3D and 5D, respectively. \par 
 
  	 We then perform  the least-squared fits to the Monte Carlo data via the ansatz Eq.~\eqref{eq:QN_ansatz}. We first discuss the $d=5$ case.  If we include only one correction term $b_1$ in the ansatz, the fits give $Q_0 = 1.569(3)$, in perfect agreement with $\pi/2$, and $y_1 = -0.56(9)$. If we fix $y_1=-1/2$, then we get the consistent estimate $1.5709(6)$.  Consistent estimates are also obtained if the $b_2$ term is added to the ansatz, with $y_2$ fixed at $-2$. We finally conclude the estimate $Q_0 = 1.569(5)$ by considering the systematic error from using different ansatz.  
  	 A similar procedure has been done for $d=2$, and we get the estimate  $Q_0 = 1.301\,0(5)$ and  $y_1 =-1.4(2)$. For $d=3$, fixing $b_2 = 0$ and leaving $Q_0,y_1,b_1$ free cannot obtain stable results. We then fix $y_2 = -2$ and leave $Q_0, b_1, b_2, y_1$ free, and we get $Q_0=1.4072(5)$, $y_1=-0.9(3)$. Fixing $y_1=-1$ also leads consistent estimate of $Q_0$ and gives $b_1=-0.067(8)$, $b_2=0.90(8)$. Finally, we estimate $Q_0 = 1.407\,2(8)$.  Details of fitting are summarized in Table~\ref{tab:fit_QN}. \par  
  \begin{table}[t]
	\centering
	\begin{tabular}{|c|l|cccccc|}
		\hline 
		$d$	&$L_{\rm min}$  	&$Q_0$ 	&$b_1$  &$b_2$ 	&$y_1$ 	&$y_2$ 	& $\chi^2/{\rm DF}$ 	\\
		\hline 
		&8     &1.3008(1) 	&0.54(1)   	&-     &-1.33(1)   &-		&2.8/7\\ 
		&16    &1.3008(2) 	&0.56(7)   	&-     &-1.34(5)   &-		&2.7/6\\ 
		2	&32    &1.3009(3) 	&0.6(4)    	&-     &-1.4(2)    &-		&2.7/5\\ 
		\cline{2-8}
		&32    &1.3010(1) 	&0.94(3)   	&-     &-3/2       &-	 	&3.1/6\\ 
		&64    &1.3010(2) 	&1.0(2)    	&-     &-3/2       &-	 	&2.9/5\\ 
		\cline{2-8}
		&8     &1.3010(1) 	&1.08(3)   	&-0.88(8)  	&-3/2   	&-2    	&3.6/7\\ 
		&16    &1.3009(2) 	&1.2(1)    	&-1.3(4)   	&-3/2   	&-2    	&2.8/6\\ 
		\hline
		&6     &1.4071(3) 	&-0.06(3)  	&0.88(7)   	&-1.0(2)   	&-2   &2.4/7\\ 
		&8     &1.4072(5) 	&-0.05(4)  	&0.8(1)    	&-0.9(3)   	&-2   &2.3/6\\  
		\cline{2-8}
		3   &6     &1.4071(1) 	&-0.065(3) 	&0.89(1)   	&-1        	&-2   &2.4/8\\ 
		&8     &1.4071(1) 	&-0.065(4) 	&0.88(3)   	&-1        	&-2   &2.4/7\\ 
		&12    &1.4071(2) 	&-0.067(8) 	&0.90(8)   	&-1        	&-2   &2.3/6\\ 
		\hline 
		&12    &1.571(1)     &-0.108(4)    &-             &-0.50(3)  		&-   	  &5.5/4\\ 
		&16    &1.568(2)     &-0.12(1)     &-             &-0.58(7)  		&-   	  &3.6/3\\ 
		\cline{2-8}
		5   &12    &1.5710(2)    &-0.1080(9)	  &-              &-1/2         	&-        &5.5/5\\ 
		&16    &1.5712(3)    &-0.109(1) 	  &-              &-1/2         	&-        &5.0/4\\ 
		&24    &1.5709(6)    &-0.107(3) 	  &-              &-1/2         	&-        &4.7/3\\ 
		\cline{2-8}
		&12    &1.5712(5)  	&-0.109(3) 	  &0.02(6)       &-1/2              &-2       &5.4/4\\ 
		&16    &1.5704(9)  	&-0.103(7) 	  &-0.2(2)       &-1/2              &-2       &4.2/3\\ 
		\hline
	\end{tabular} 
	\caption{The fitting result of $Q_N$ for SAW with $d=2,3,5$.}
	\label{tab:fit_QN} 
\end{table}

\begin{figure}[H]
  	\centering
  	\includegraphics[scale=0.70]{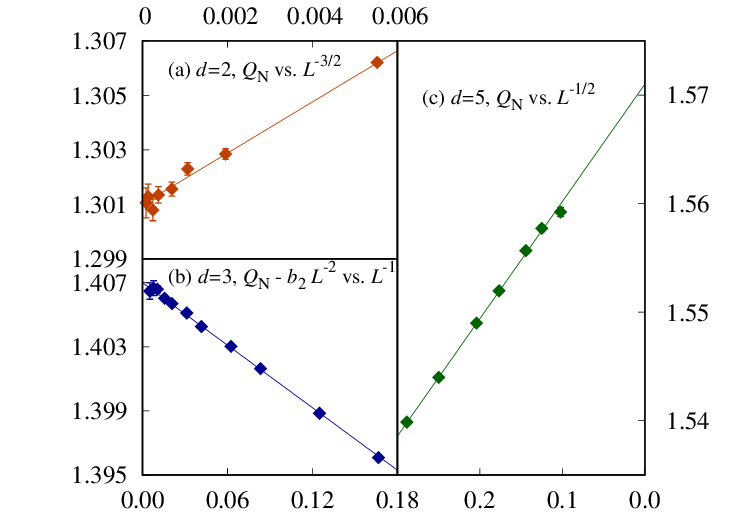}
  	\caption{Plot of the critical Binder cumulant $Q_N$ for various dimensions (a) $d=2$, (b) $d=3$ and (c) $d=5$.}
  	\label{fig:qn_d235}
  \end{figure}

  	 In Fig.~\ref{fig:qn_d235}, we plot $Q_N$ versus $L^{y_1}$ for $d=2,5$, where $y_1$ takes the value $-3/2, -1/2$ respectively.  It shows that the asymptotic values of $Q_N$ are consistent with our estimates. For $d=3$, since $b_2$ is much larger than $b_1$, to clearly show the leading correction term $b_1L^{y_1}$, we subtract $b_2L^{y_2}$ from $Q_N$.  The asymptotic value of  $Q_N - b_2 L^{y_2}$ in the $L \to \infty $ limit is close to 1.407, which is consistent with our estimate. 
  \par  
   
\begin{figure}[H]
  	\centering
  	\includegraphics[scale=0.70]{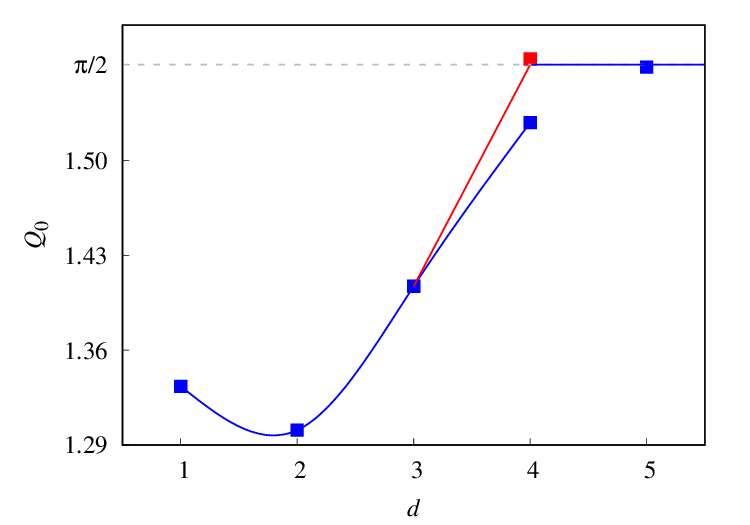}
  	\caption{Plot of the critical Binder cumulant $Q_0$ for $d=1,2,3,4,5$. The gray dashed line shows the value $\pi/2$. Both the two possible scenarios, that $Q_0$ is consistent with $\pi/2$ (red) or less than $\pi/2$ (blue), are shown for the 4D case. The horizontal blue line indicates that $Q_0$ takes the complete-graph value $\pi/2$ for $d>4$. }
  	\label{fig:q0_d12345}
  \end{figure}

  	For $d=1$, the critical point is trivially at  $\zc= 1$ since the number of $N$-step SAWs rooted at an arbitrarily fixed point is $c_N=2$ for all $N$.
  	Thus, for SAW on a length-$L$ cycle, the critical mean walk length is 
  		\begin{equation}
  		 	\langle \scrN \rangle = \frac{\sum_{N=0}^{L-1} N c_N z_c^N}{\sum_{N=0}^{L-1} c_N z_c^N}=\frac{\sum_{N=1}^{L-1} 2N}{1 + \sum_{N=1}^{L-1} 2}= \frac{L(L-1)}{2L-1}, \nonumber 
  		\end{equation}
  	and its second moment is 
  	\begin{equation}
  		 \langle \scrN^2 \rangle = \frac{\sum_{N=0}^{L-1} N^2 c_N z_c^N}{\sum_{N=0}^{L-1} c_N z_c^N}= \frac{L(L-1)}{3},      \nonumber 
  	\end{equation} 
  	and the Binder cumulant
  	      \begin{equation}
  	      	Q_N = \frac{\langle \scrN^2 \rangle }{\langle \scrN\rangle^2} = \frac{(2-1/L)^2}{3(1-1/L)}.
  	      \end{equation}
        In the $L \to \infty$ limit, we get $Q_N = 4/3$. 
  	  In Fig.~\ref{fig:q0_d12345}, we plot the estimates of $Q_0$ for various $d$. For $1\le d \le 3$, we connect the data points with a smooth curve. For 4D, we show the two possible scenarios from our data analysis: $Q_0$ is consistent with the complete-graph value $\pi/2$ (red) or $Q_0$ is less than $\pi/2$ (blue). For $d>4$, $Q_0$ takes the complete-graph value $\pi/2$. 
  
  \par 

\bibliographystyle{apsrev4-1}
%

\end{document}